\documentclass[conference]{IEEEtran}
 \usepackage[pdftex]{graphicx}
 \usepackage{enumitem}

\ifCLASSINFOpdf
\else
\fi

\usepackage {bsymb,b2latex}

\begin{document}
%
\title{Formal Modelling, Testing and Verification of HSA Memory Models using Event-B}

\author{\IEEEauthorblockN{Ashish Darbari\\ and Iain Singleton}
\IEEEauthorblockA{Imagination Technologies\\
Kings Langley\\
Hertfordshire WD4 8LZ, UK\\
}
\and
\IEEEauthorblockN{Michael Butler\\and John Colley}
\IEEEauthorblockA{University of Southampton\\
Southampton\\ Hampshire SO17 1BJ, UK\\
}
}
\maketitle

\begin{abstract}
The Heterogeneous System Architecture (HSA) Foundation has produced the HSA Platform System Architecture Specification that goes a long way towards addressing the need for a clear and consistent method for specifying weakly consistent memory. A weakly-consistent memory model is one of the fundamental cornerstones for achieving high performance concurrency with low power on mobile platforms. HSA is specified using a natural language which makes it open to multiple ambiguous interpretations and could thereby render  bugs in implementations of it in hardware and software. In this paper, we present a formal model of HSA which can be used in the development and verification of both concurrent software applications as well as in the development and verification of the HSA-compliant platform itself. We use the Event-B language to build a provably correct hierarchy of models from the most abstract to a detailed refinement of HSA that is close to implementation level. Our memory models are general in the sense that they represent arbitrary numbers of masters, programs and instruction interleavings and we reason about such general models using theorem proving.  By using the Rodin tool for Event-B we are able to seamlessly model and verify the entire hierarchy of models using proofs to establish that each refinement is correct. We also define an automated validation method that allows us to check baseline compliance of the model against a suite of published HSA litmus test cases.
Once we have completed the model validation we develop a coverage driven method to extract a richer set of test cases from the formal Event-B model and a user specified coverage model. These tests are then used for extensive regression testing of hardware and software systems. We believe our methodology of refinement based formal modelling, baseline compliance testing of the formal model and coverage driven test extraction using a single language of Event-B language and the Rodin tool is a completely new way of addressing a profoundly important challenge facing the design and verification of low-power of multi-core systems.
\end{abstract}


%
\IEEEpeerreviewmaketitle

\section{Introduction}
Weakly-consistent memory \cite{dubois1986memory} is one of the fundamental cornerstones for achieving high performance concurrency with low power on mobile platforms.  Although the buffering mechanisms that underly weakly consistent memory models are relatively straightforward to understand, developing a high-level, natural language specification that represents the behaviours of all implementations is very difficult.  This presents a challenge to both software engineers, who wish to develop efficient, race-free concurrent programs and for platform engineers who wish to develop systems that conform to a particular memory model.

The Heterogeneous System Architecture (HSA) Foundation is a not-for-profit industry standards body which has produced the HSA Platform System Architecture Specification \cite{HSAPSAS} that goes a long way towards addressing the need for a clear and consistent method for specifying weakly consistent memory, but since it is still in natural language  there is still a need for a more formal representation.  In this paper, we describe how the HSA specification and the well-founded terminology and concepts it describes, is used to develop a formal model in Event-B \cite{AbrialBHHMV10} which can be used in the development and verification of both concurrent software applications as well as in the development and verification of the HSA-compliant platform itself, the specific configuration of masters, slaves and interconnect.

We use the Event-B~\cite{AbrialBHHMV10}  language to build a provably correct hierarchy of models from the most abstract to a detailed refinement of HSA that is close to implementation level. Our memory models are general in the sense that they represent arbitrary numbers of masters, programs and instruction interleavings and we reason about such general models using theorem proving.  By using the Rodin tool for Event-B we are able to seamlessly model and verify the entire hierarchy of models using proofs to establish that each refinement is correct. We also define an automated validation method that allows us check baseline compliance of the model against a suite of published HSA litmus test cases. The validation method involves refining the general models to represent litmus tests and uses model checking to verify that all possible interleaving of a litmus test achieve the expected outcome. Once we have completed the model validation we develop a coverage driven method to extract a richer set of test cases from the formal Event-B model and a user specified coverage model. These tests are then used for extensive regression testing of hardware and software systems. We believe our methodology of refinement based formal modelling, baseline compliance testing of the formal model and coverage driven test extraction using a single language of Event-B language and the Rodin tool is a completely new way of addressing a profoundly important challenge facing the design and verification of low-power of multi-core systems.

\section{Event-B based Formal Modelling of HSA}

Event-B is a proof-based modelling language and method  that enables the systematic development of specifications using a formal notion of refinement. The Rodin platform \cite{AbrialBHHMV10} is the Eclipse-based IDE that provides automated support for Event-B modelling, refinement and mathematical proof, model-checking and model-based test generation.  

Our approach starts at the abstract level, focusing on issue and observation of memory instructions.    In subsequent refinements, we introduce \emph{set-theoretic relations} on the memory accesses which constrain the order in which they can be issued or observed. 
At each refinement level we are able to verify  properties formally at their appropriate level of abstraction and detailed instruction sets are introduced in correctness-preserving refinement steps. 
This refinement-based approach helps to manage complexity of the modelling and verification.

Using this modelling approach, we build a formal memory model that represents an arbitrary set of masters running an arbitrary set of program threads.  We then use further Event-B refinements to constrain the model to represent a fixed set of masters, each executing a specific program thread.  In this way, we can represent a published \emph{HSA litmus test}, together with a property or set of properties which specify the desired \emph{outcome} of the test. 
We then use a combination of \emph{automatic theorem proving, model checking and simulation} to verify and validate the formalised memory models against the formalised litmus tests.  
Once we have completed the model validation we develop a coverage driven method to extract a richer set of test cases from the formal Event-B model and a user specified coverage model. These tests can then be used for extensive regression testing of hardware and software systems. 

\begin{figure}[!htb]
  \centering
  \includegraphics[width=6cm, height=8cm]{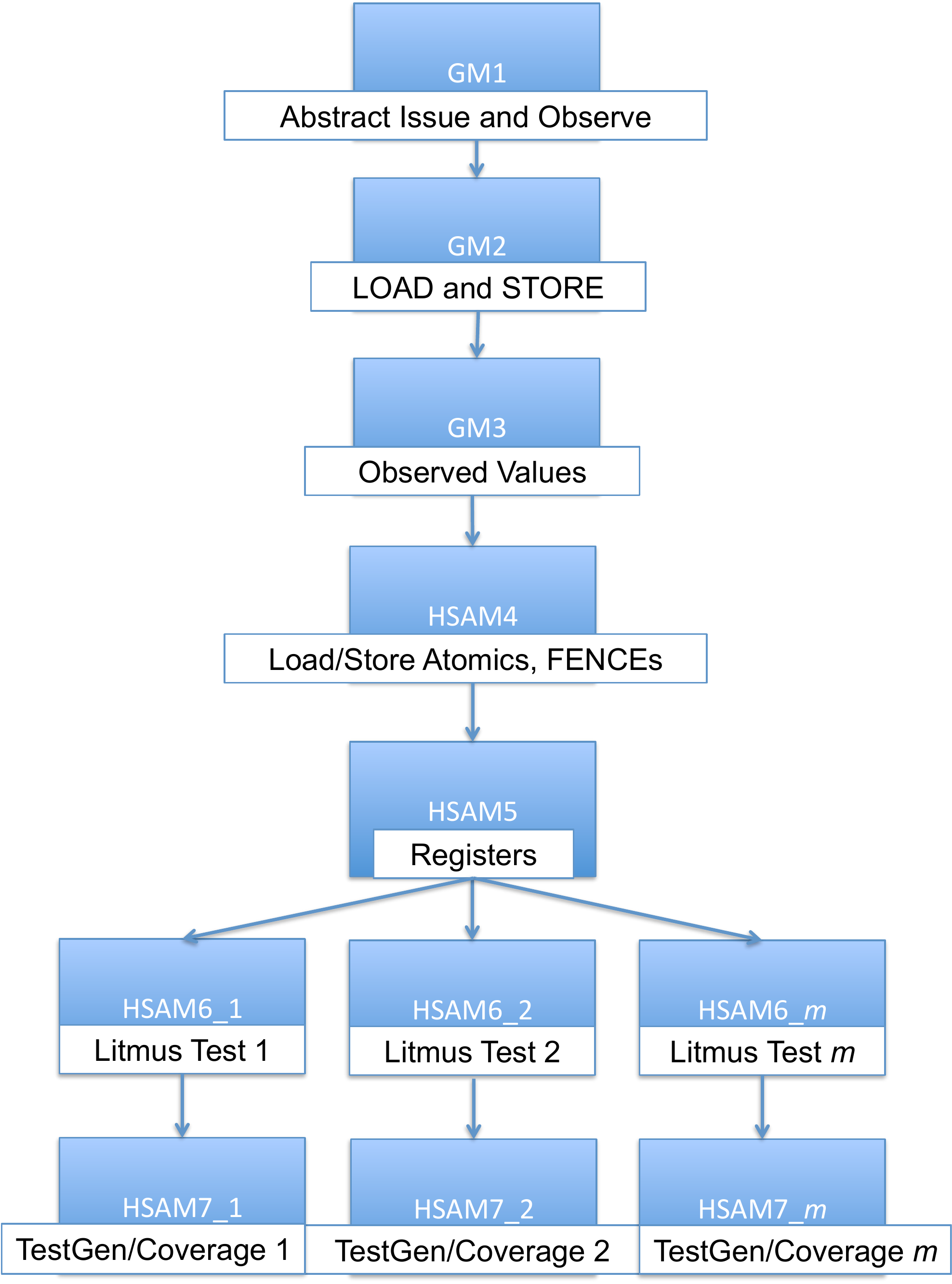}   
  \caption{Model Refinement Hierarchy}
  \label{Figure:figHIERFULL} 
\end{figure}

In Figure \ref{Figure:figHIERFULL} below we show the Event-B model refinement hierarchy. 
The abstract generic models are denoted by \emph{GMn}, where \emph{n} ranges from 1 to 3 whilst the HSA-specific  models are denoted by \emph{HSAMn[\_m]} where \emph{n} ranges from 4 upwards and \emph{\_m} represents a litmus test at the concrete level. 
The abstract models, \emph{GM1} to \emph{GM3}, are generic and can form the foundation for modelling, for instance,  Sequential Consistency (SC) \cite{lamport1979make}, Total Store Order (TSO) \cite{sewell2010x86} or ARM \cite{chong2008reasoning} memory models.  In \emph{GM1} we define the fundamental memory accesses with the  events \emph{Issue} and \emph{Observe}, in \emph{GM2} we differentiate between \emph{LOAD} and \emph{STORE} memory accesses and in \emph{GM3} we model the \emph{observation} of the LOAD and STORE values.  In the next refinement, \emph{HSAM4}, we model the ordering rules for HSA atomics and FENCEs and in refinement \emph{HSAM5} we model the registers generically.
We then introduce a series of refinements, \emph{HSAM6[1 .. m]} to model a series of \emph{m} concurrent programs, the litmus tests, each running on a fixed set of masters and use the ProB model checker \cite{LeuschelB08} to validate the model. Each of these models is then further refined, \emph{HSAM7[1 .. m]} to introduce the functional coverage metric for coverage measurement and test generation.

Once we have completed the model validation and have a set of litmus tests which cover the model, we relax the constraints on the litmus test models \emph{HSAM7[1 .. m]} so that instead of each representing a single, fixed concurrent program, each model represents a class of concurrent, synchronised programs, based on the litmus test, from which we can generate tests automatically that can be used in the development and verification of the HSA-compliant platform itself.

%
%
%
%

In summary, the contributions of this paper are:
\begin{itemize}
\item a hierarchy of formalised and verified memory models with varying ordering constraints
\item a method for validating models against known litmus tests through model-checking
\item a method for automatic generation of tests from a constrained model for regression testing of an HSA-compliant platform.
\end{itemize}

\section{Understanding HSA Ordering}

We develop a generalised approach to weakly consistent memory modelling, based on HSA terminology, which provides the building blocks with which we can model weak memory semantics with a single, consistent modelling framework. We also provide at the concrete level a generalised notation for representing litmus tests and the invariants associated with those tests.


We begin with a definition of \emph{load} and \emph{store} from the HSA reference manual \cite{HSAPSAS}. We focus our modelling around LOAD and STORE operations that may be issued and observed by different units of execution.  

\begin{itemize}

\item For primitives of type \emph{store}, visibility to unit of execution A of a store operation X is when the data of store X is available to \emph{loads} from unit of execution A.

\item For primitives of type \emph{load}, visibility is when a load gets the data that the unit of execution will put in the \emph{register}.
\end{itemize}

We shall use the HSA definition as the basis for formalising the specification of \emph{loads} and \emph{stores}.

\begin{figure}[!htb]
  \centering
  \includegraphics[width=8cm, height=1.3cm]{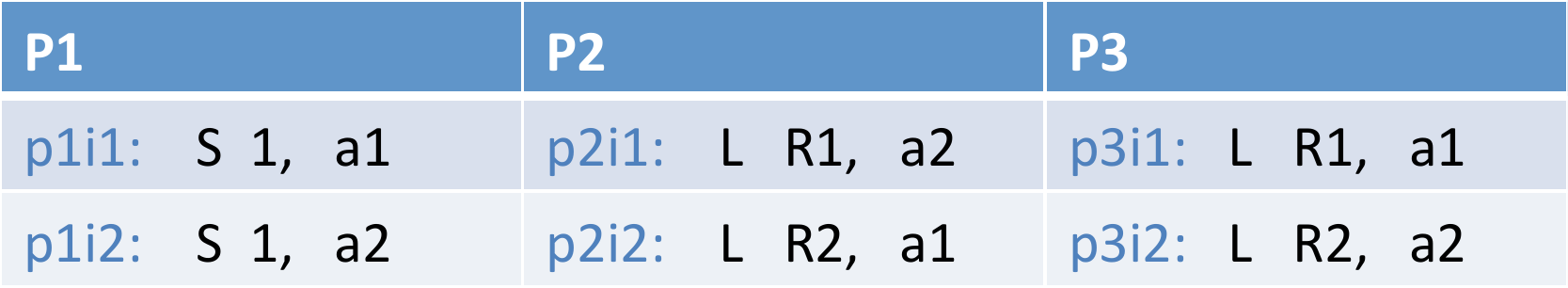}   
  \caption{Litmus test}
  \label{Figure:figlitmusREV} 
\end{figure}

\subsection{Observation Ordering}

Without considering synchronisation, based on the HSA definition above, we define observation ordering, using the litmus test of Figure \ref{Figure:figlitmusREV} to illustrate the definitions.  We assume that the values at addresses \emph{a1} and \emph{a2} are initially set to \emph{0}

\begin{itemize}
\item A STORE is observed by any master when a subsequent LOAD issued by that master would result in the LOAD returning the value associated with the STORE
\begin{itemize}
\item \emph{The STOREs p1i1 and p1i2 are observed by masters P1, P2, P3 in some order and these STORE observations with respect to the LOAD observations determine the value which the  LOADs observe. }
\end{itemize}
\item A LOAD is observed by the master issuing the LOAD when the LOAD returns the value associated with the last observed STORE to the LOAD's address
\begin{itemize}
\item \emph{The LOADs p2i1 and p2i2 are observed by P2 in some order; similary for the P3 LOADS.}

\end{itemize}
\end{itemize}

Note that this definition of observation ordering does not refer to values in registers.  We use, instead, the more abstract concept of a LOAD \emph{returning} a value.  We also use the term \emph{observe}  \cite{Shen99commit-reconcilefences} rather than the HSA \emph{visibility}.

For \emph{Synchronisation}, the ordering of STORE and LOAD observations is constrained
\begin{enumerate}
\item An ordering is established between LOAD and STORE observations when those LOADs and STOREs are issued by a single master (e.g., program order). In Figure \ref{Figure:figlitmusREV3}, the arrow indicates the ordering of the LOADs on P2 and P3.

\begin{figure}[!htb]
  \centering
  \includegraphics[width=8cm, height=1.6cm]{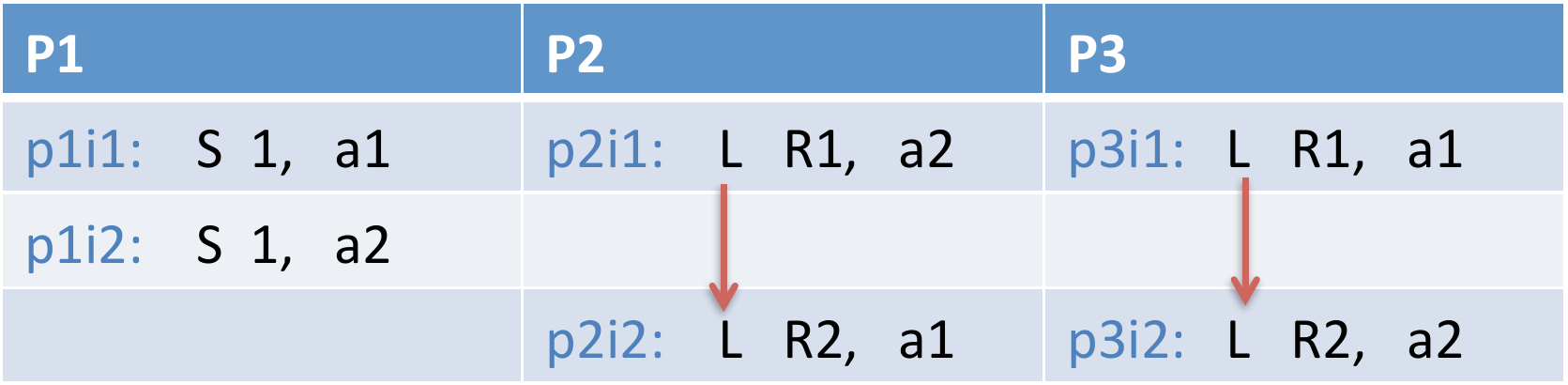}   
  \caption{Ordering between LOADs on the the same master}
  \label{Figure:figlitmusREV3} 
\end{figure}

\item An ordering is established between LOAD observations on different masters to the same address as shown in Figure \ref{Figure:figlitmusREV4}.

\begin{figure}[!htb]
  \centering
  \includegraphics[width=8cm, height=1.6cm]{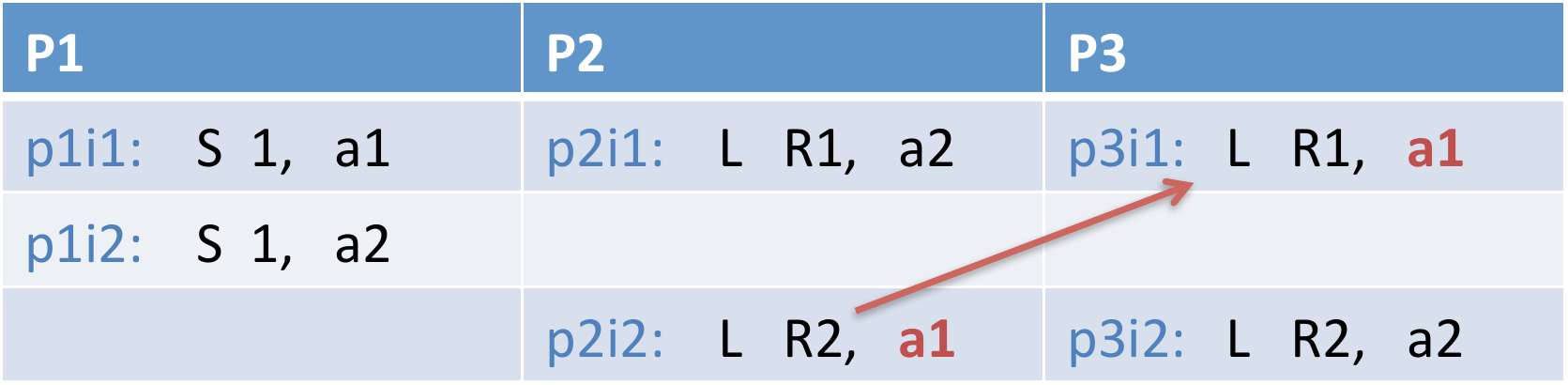}    
  \caption{Ordering between LOADs on different masters}
  \label{Figure:figlitmusREV4} 
\end{figure}

\item An ordering is established between a STORE observation by a given master and an observation of a LOAD by a different master for the same address as shown by the labelled arrow in Figure \ref{Figure:figlitmusREV5}. \emph{P2} observes the STORE \emph{p1i2} before it observes the LOAD \emph{p2i1}.

\begin{figure}[!htb]
  \centering
  \includegraphics[width=8cm, height=1.6cm]{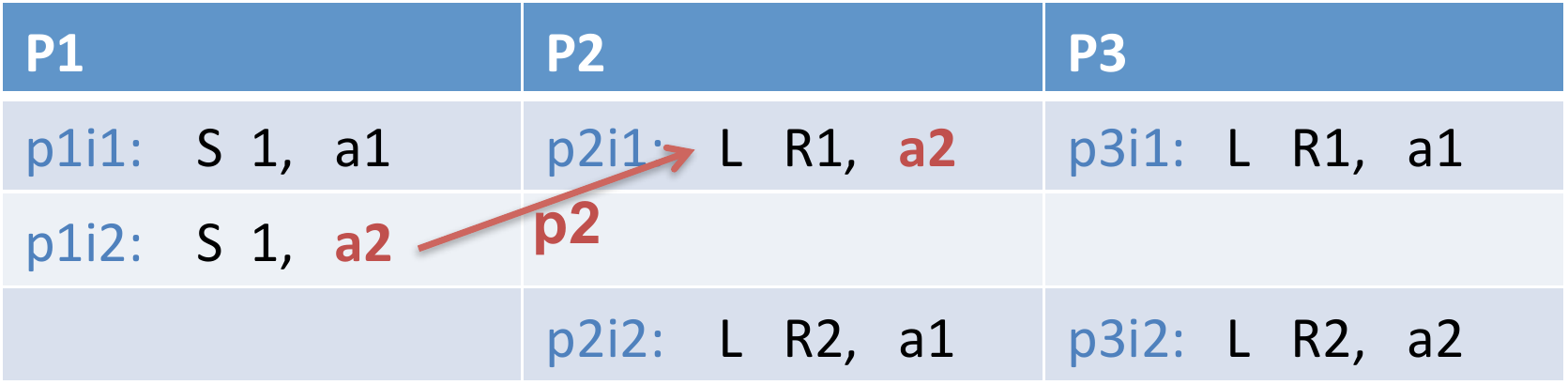}  
  \caption{Ordering between STOREs and LOADs on different masters}
  \label{Figure:figlitmusREV5} 
\end{figure}

\end{enumerate}

With these three orderings in place, the following outcome is not possible:- 

(p2:R1 = 1 and p2:R2 = 0 and p3:R1 = 1 and p3:R2 = 0) 

For this outcome to occur would mean that

\begin{figure}[!htb]
  \centering
  \includegraphics[width=8cm, height=1.6cm]{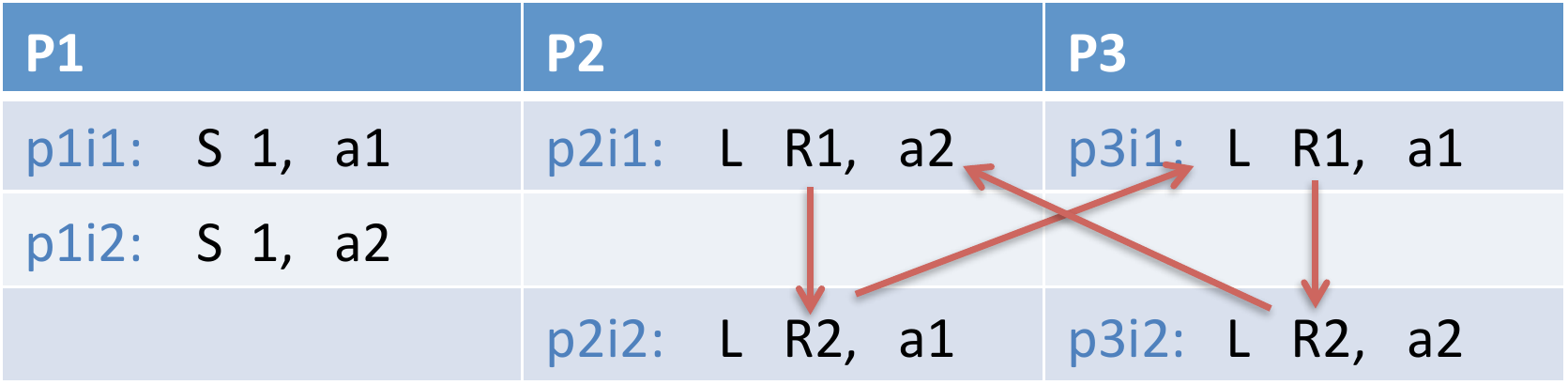}  
  \caption{Invalid ordering}
  \label{Figure:figlitmusREV6}  
\end{figure}

\begin{itemize}
\item p2i1 and p2i2 are observed in program order
\item P2 must observe the LOAD p2i2 before P3 observes the LOAD p3i1
\item P3 must observe the LOAD p3i2 before P2 observes the LOAD p2i1
\item This ordering is invalid (cyclic), therefore the assumed outcome is not possible, as shown in Figure \ref{Figure:figlitmusREV6}.
\end{itemize}

We now  describe these three synchronisation orderings, using the definitions of \emph{Program Order (po)}, \emph{Coherent Order (co)} and \emph{Happens Before Order (hb)}, as defined in the HSA specification.  
For \emph{po}, there is a total sequential order on operations within a single unit of execution, for \emph{co}, there is a total apparent order on all synchronisation operations consistent with \emph{po} and  \emph{hb} must be consistent with each \emph{co} and with all \emph{SC} orders.

We illustrate these orderings using the same litmus test.

\subsection{LOADs and STOREs Observed in Program Order (po)}

\emph{Memory accesses from the same master are observed in the order they are issued by all masters.}
\\
\\
Figure \ref{Figure:figlitmusREV3} illustrates \emph{po} on masters \emph{P2} and \emph{P3}.

%


\subsection{Coherent Observation of LOADs/STOREs per Master (co)}

\emph{If a master observes a STORE followed by a LOAD for the same memory address, then the value it reads in the LOAD is the value defined in the STORE.}
\\
\\
Figure \ref{Figure:figlitmusREV4} illustrates \emph{co} between masters \emph{P2} and \emph{P3}.

%
%
%
%
%
%
%


\subsection{Happens Before ordering of Load observations (hb)}

\begin{figure}[!htb]
  \centering
  \includegraphics[width=8cm, height=1.6cm]{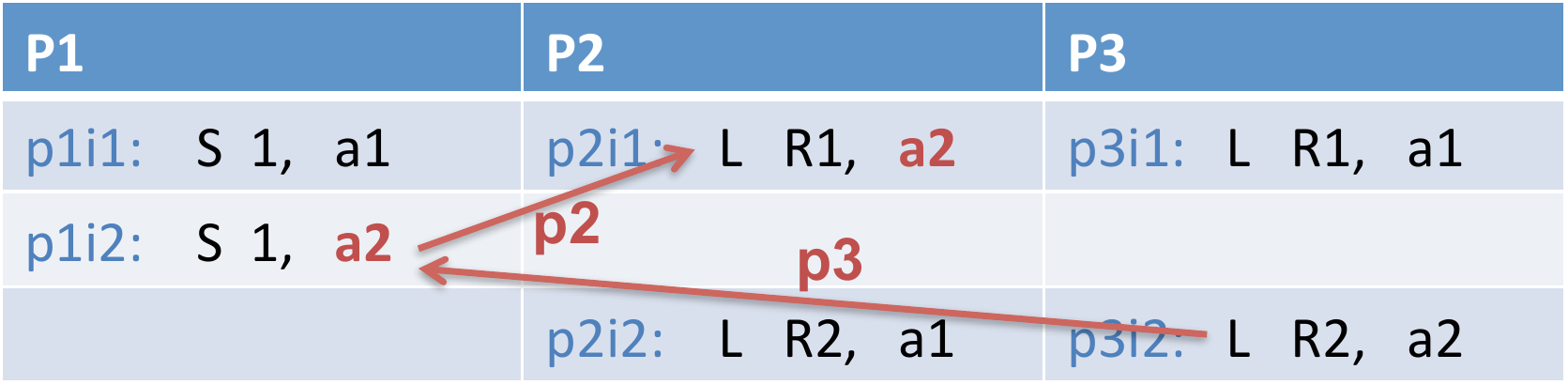}  
  \caption{Happens Before (hb)}
  \label{Figure:figHBlitmus} 
\end{figure}

Finally, we introduce the \emph{Happens Before} ordering.
\emph{This is an ordering between loads  defined by a combination of coherent orderings on loads and stores to the same address, i.e., if a load L1 from an address A is observed before a store S1 to A and load L2 is observed after S1 by some master, then L1 is observed before L2 by all masters.}
(See \ref{Figure:figHBlitmus}.)

%
%
%
%
%

\subsection{Synchronisation Mechanisms for Weakly Consistent Memory }

Using these notions of observation ordering, \emph{po}, \emph{co}, and \emph{hb}, we can now define the mechanisms by which synchronisation can be achieved.  First, we look at the case of \emph{No Synchronisation} for weakly consistent memory

\subsubsection{No Synchronisation}

If no fences or specific synchronising instructions are used, then the only ordering is \emph{co}, as shown in Figure \ref{Figure:figlitmusREV}.
No synchronisation means that stores are observed by different masters in different order.


\subsubsection{Synchronising Memory Accesses (Single Memory Location)}

\begin{figure}[!htb]
  \centering
  \includegraphics[width=8cm, height=1.3cm]{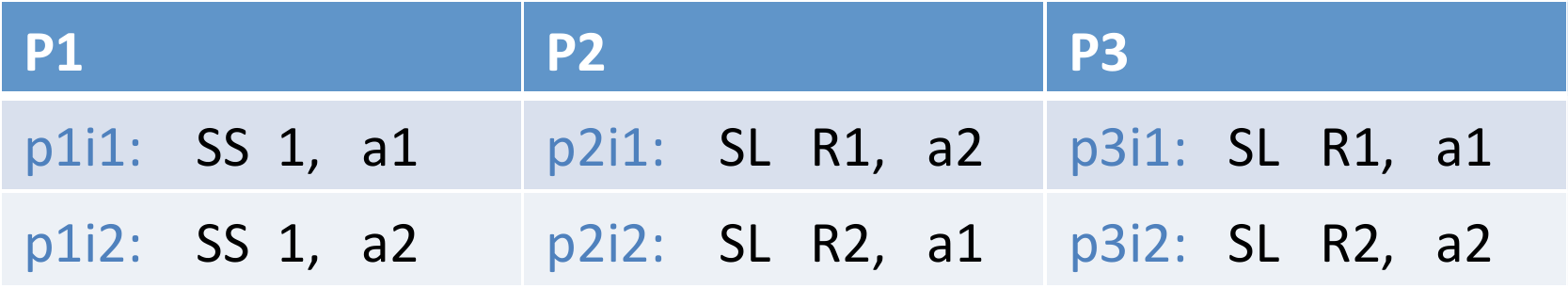}  
  \caption{Synchronising Memory Accesses}
  \label{Figure:figATOM} 
\end{figure}

HSA supports atomic synchronizing operations with \emph{acquire} and \emph{release} semantics and are, by definition, \emph{Sequentially  Consistent}.  The operations apply to a single memory location.  These synchronising LOADs and STOREs, represented as \emph{SL} and \emph{SS} in Figure \ref{Figure:figATOM}, result in \emph{po}, \emph{co} and \emph{hb} observation orderings. \emph{SL} and \emph{SS} instructions impose \emph{po} and \emph{hb} ordering.

\subsubsection{Fences (Multiple Memory Locations in the same Scope)}

\begin{figure}[!htb]
  \centering
  \includegraphics[width=8cm, height=1.6cm]{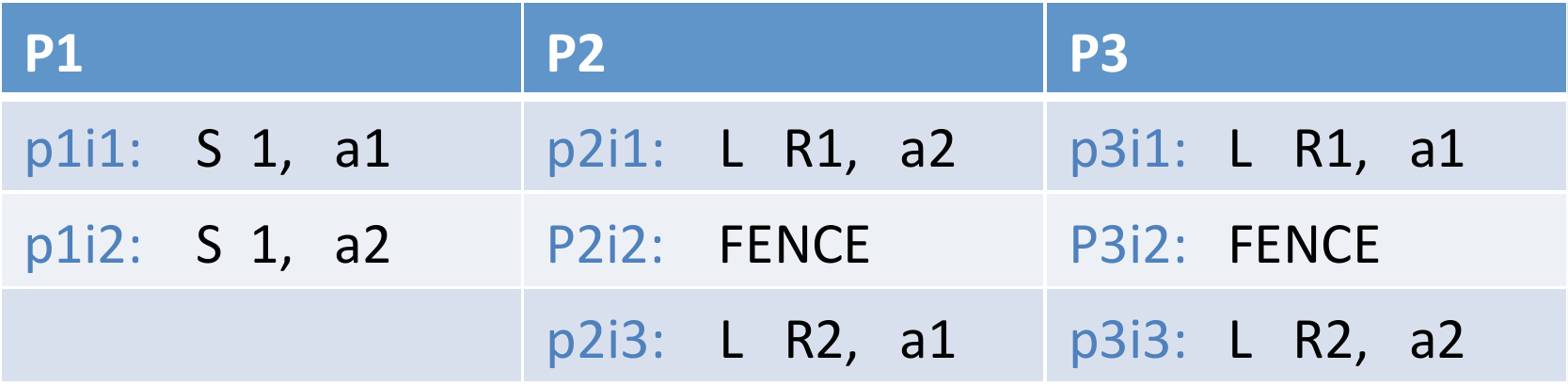}  
  \caption{Fence Synchronisation}
  \label{Figure:figHSAFL} 
\end{figure}

Instead of using synchronising memory accesses, we can introduce a \emph{FENCE} instruction between the LOADs on \emph{P3} and \emph{P4}  as shown in  Figure \ref{Figure:figHSAFL}.  The introduction of the FENCE also results in \emph{po}, \emph{co} and \emph{hb} observation orderings. Whereas the LOAD and STORE atomics only apply to a single memory location, the FENCE applies to all memory locations within the same defined scope, as described in the HSA specification.


\section{Modelling Weak Memory with Event-B}

We begin with an abstract, highly nondeterministic model of memory accesses.  The abstract Event-B context defines a set of transactions, \emph{TRN} and a subset of the transactions which are memory accesses, \emph{MEMACCESS}.

\begin{description}
		\nItem{ axm1 }{ MEMACCESS \subseteq  TRN }
\end{description}

We assume a set of masters that issue transactions on a shared memory space and the effect of those transactions is observed some time after they are issued.  Key to defining weak memory models is the relative ordering, locally and globally, of transaction issues and observations.  For this reason, we start our modelling with the key events in weak memory systems, Issue transaction and Observe transaction.

We therefore define an abstract Event-B machine, \emph{GM1},  with two events, \emph{Issue} and \emph{Observe}, which operate on  memory accesses.

\begin{description}
    \nItem{ inv1 }{ issued \subseteq  MEMACCESS }
    \nItem{ inv2 }{ observed \subseteq  issued }
\end{description}
	
\begin{description}
	\EVT {Issue}
		\begin{description}
		\AnyPrm
			\begin{description}
			\Item{ma }
			\end{description}
		\WhereGrd
			\begin{description}
			\nItem{ grd1 }{ ma \in  MEMACCESS }
			\nItem{ grd2 }{ ma \notin  issued }
			
			\end{description}
		\ThenAct
			\begin{description}
			\nItem{ act1 }{ issued :=  issued \bunion  \{ ma\}  }
			\end{description}
		\EndAct
		\end{description}
\end{description}

		
\begin{description}
	\EVT {Observe}
		\begin{description}
		\AnyPrm
			\begin{description}
			\Item{ma }
			\end{description}
		\WhereGrd
			\begin{description}
			\nItem{ grd1 }{ ma \in  issued }
			\end{description}
		\ThenAct
			\begin{description}
			\nItem{ act1 }{ observed :=  observed \bunion  \{ ma\}  }
			\end{description}
		\EndAct
		\end{description}
\end{description}

The variable \emph{issued}, invariant \emph{inv1}, initialised to the empty set, records the issuing of each memory access by the event \emph{Issue}.  The parameter \emph{ma} of the event \emph{Issue} represents a memory access, the first guard, \emph{grd1}, ensures that the memory access cannot be re-issued if it has already been issued and \emph{ma} is added to the \emph{issued} set by the action \emph{act1}. 

The variable \emph{observed}, invariant \emph{inv2} is defined as a subset of the memory accesses that have been issued.  If the set \emph{issued} is not empty, the event \emph{Observe} chooses an issued memory access non-deterministically from the set and the action \emph{act1} adds it to the set \emph{observed}.  At this early stage there is no notion of a program.  Memory accesses are issued in some arbitrary order, and once issued they can be observed \emph{repeatedly} in some arbitrary order.

We then refine the abstract model to differentiate between \emph{LOAD} and \emph{STORE} accesses,

\begin{description}
		\nItem{ axm2 }{ partition(MEMACCESS, STORE, LOAD) }
\end{description}

and refine the abstract events \emph{Issue} and \emph{Observe}.

\begin{itemize}
\item New events \emph{IssueLoad} and \emph{IssueStore} refine \emph{Issue}
\item New events \emph{ObserveLoad} and \emph{ObserveStore} refine \emph{Observe}
\end{itemize}

In the next refinement, we introduce an abstraction of the memory architecture, which makes reasoning about weak memory consistency tractable\cite{Shen99commit-reconcilefences}. Each master keeps track individually of its last observed STORE value for each memory location. When a LOAD is observed, the last observed value for that master is written to the master's target register.

An observation ordering is then established between LOADs and STOREs, depending on address, control and data dependencies and the positioning of fences.

\begin{figure}[!htb]
  \centering
  \includegraphics[width=6.5cm, height=3.5cm]{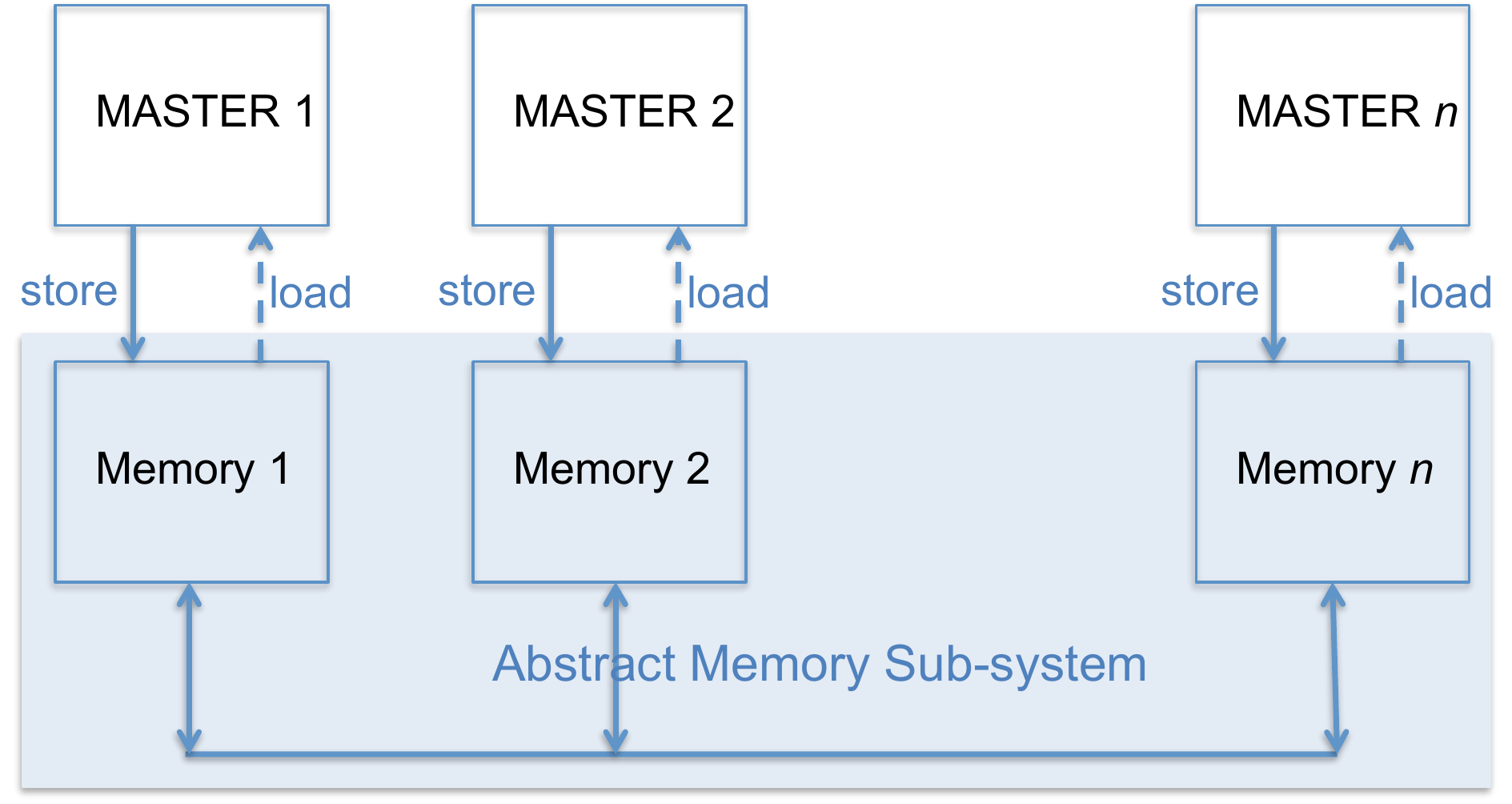}
  \caption{Abstract Memory Architecture}
  \label{Figure:figARM} 
\end{figure}

The abstract architecture is shown in Figure \ref{Figure:figARM}, where each \textit{Memory i} represents the last observed STORE value of each memory location for each individual \textit{Master i}.

This architecture is represented in Event-B with the variable \emph{lov}.

\begin{description}
		\nItem{ inv1 }{ lov \in  MASTER \tfun  (ADDR \tfun  VALUE) }
\end{description}

which maintains the \emph{last observed value}.

We also introduce the variable \emph{observers}, which associates each memory access with the masters which have observed that access.

\begin{description}
		\nItem{ inv2 }{ observers \in  MEMACCESS \pfun  \pow(MASTER) }		
\end{description}

For each LOAD or STORE that is issued, \emph{observers(s)} is initialised to the empty set.

The refined event \emph{ObserveStore} introduces a new guard, \emph{grd3}, to ensure that this master, \emph{m}, has not already observed this store, \emph{s}, an action, \emph{act2},  which updates the last observed value for \emph{m}, using the \emph{relational override} operator, and an action, \emph{act3}, which adds \emph{m} to the observers of \emph{s}.

\begin{description}
	\EVT {ObserveStore}
	\REF {ObserveStore}
		\begin{description}
		\AnyPrm
			\begin{description}
			\Item{s, m }
			\end{description}
		\WhereGrd
			\begin{description}
			\nItem{ grd1 }{ s \in  issued }
			\nItem{ grd2 }{ s \in  STORE }
			\nItem{ grd3 }{ m \notin  observers(s) }
			\end{description}
		\ThenAct
			\begin{description}
			\nItem{ act1 }{ observed :=  observed \bunion  \{ s\}  }
			\nItem{ act2 }{ lov(m) :=  lov(m) \ovl  \{ addr(s) \mapsto  val(s)\}  }
			\nItem{ act3 }{ observers(s) :=  observers(s) \bunion  \{ m\}  }
			\end{description}
		\EndAct
		\end{description}
\end{description}

We now have a generic, abstract model which can now be refined to implement HSA Atomics and Fences.

\subsection{Modelling HSA Fence Synchronisation}

Using the definitions of \emph{po}, \emph{co} and \emph{hb} from the HSA specification, we are  able to model the observation orderings necessary to implement fence synchronisation.  We introduce the variable \emph{ahead} to represent the accesses that are ahead of the fence and the variable \emph{after} to represent the set of LOADs that are observed after a given STORE.
Our refinement distinguishes multiple cases of LOAD observation, each of which is represented by an event that refines our abstract \emph{ObserveLoad} event.
First, we specify the LOAD observations in the presence of a FENCE. There are two cases to address, the load happening before a store and happening after a store.

\emph{ObserveLoadHappensBeforeWithFence:}
If the master that issued the LOAD has issued a FENCE and if the LOAD is not ahead of the FENCE in program order, then all other memory accesses ahead of the FENCE in program order must have been observed (guard  \emph{grd7} below).
This LOAD is observed before a corresponding STORE to the same location (guards \emph{grd9, grd10)} and no other LOAD has observed that STORE  and there is therefore no LOAD \emph{after} the STORE (guard \emph{grd11)}.

\begin{itemize}
\item LOADs are observed in \emph{Program Order(po)}

\item LOAD is observed before a corresponding STORE

\item No other LOAD has been observed after the STORE

\item po + co + hb
\end{itemize}
The event is specified as follows:
\begin{description}

\EVT {ObserveLoadHappensBeforeWithFence}
	\EXTD {ObserveLoad}
		\begin{description}
		\AnyPrm
			\begin{description}
			\ItemX{l, m, f, s }
			\end{description}
		\WhereGrd
			\begin{description}
			\nItemX{ grd1 }{ l \in  issued } 
			\nItemX{ grd2 }{ l \in  LOAD }
			\nItemX{ grd3 }{ m \notin  observers(l) }
			\nItemX{ grd4 }{ m = issuer(l) }
			\nItemX{ grd5 }{ f \in  issuedfence }
			\nItemX{ grd6 }{ m = issuer(f) }
			\nItemX{ grd7 }{ l \notin  ahead(f) \limp \\ (\forall a\qdot a \in  ahead(f) \land  a \in  dom(observers) \limp \\ m \in  observers(a)) }
			\nItemX{ grd8 }{ s \in  STORE }
			\nItemX{ grd9 }{ address(s) = address(l) }
			\nItemX{ grd10 }{ m \notin  observers(s)  }
			\nItemX{ grd11 }{ after(s) = \emptyset  }
			\end{description}
		\ThenAct
			\begin{description}
			\nItemX{ act1 }{ observed :=  observed \bunion  \{ l\}  }
			\nItemX{ act2 }{ observers(l) :=  observers(l) \bunion  \{ m\}  }
			\end{description}
		\EndAct
		\end{description}
\end{description}

%
%
%
%
%
%
%
%

\emph{ObserveLoadAfterStoreWithFence:}
If the master that issued the LOAD has issued a FENCE and if the LOAD is not ahead of the FENCE in program order, then all other memory accesses ahead of the FENCE in program order must have been observed \emph{(guard grd7)}.
The master has observed a corresponding STORE to the same location \emph{(guard grd11)}.

\begin{itemize}
\item In Program Order
\item LOAD is observed after a corresponding STORE
\item po + co
\end{itemize}
The event is specified as follows:
\begin{description}
	\EVT {ObserveLoadAfterStoreWithFence}
	\EXTD {ObserveLoad}
		\begin{description}
		\AnyPrm
			\begin{description}
			\ItemX{l, m, f, s }
			\end{description}
		\WhereGrd
			\begin{description}
			\nItemX{ grd1 }{ l \in  issued }
			\nItemX{ grd2 }{ l \in  LOAD }
			\nItemX{ grd3 }{ m \notin  observers(l) } 
			\nItemX{ grd4 }{ m = issuer(l) }
			\nItemX{ grd5 }{ f \in  issuedfence }
			\nItemX{ grd6 }{ m = issuer(f) }
			\nItemX{ grd7 }{ l \notin  ahead(f) \limp \\ (\forall a\qdot a \in  ahead(f) \land  a \in  dom(observers) \limp \\ m \in  observers(a)) }
			\nItemX{ grd8 }{ s \in  issued }
			\nItemX{ grd9 }{ s \in  STORE }
			\nItemX{ grd10 }{ address(s) = address(l) }
			\nItemX{ grd11 }{ m \in  observers(s) }
			\end{description}
		\ThenAct
			\begin{description}
			\nItemX{ act1 }{ observed :=  observed \bunion  \{ l\}  }
			\nItemX{ act2 }{ observers(l) :=  observers(l) \bunion  \{ m\}  }
			\nItemX{ act3 }{ after(s) :=  after(s) \bunion  \{ l\}  }
			\end{description}
		\EndAct
		\end{description}
\end{description}

%
%
%
%

Similarly, we specify the two cases of LOAD observation \emph{without} a FENCE covering loads  before and after a store.  We also specify an \emph{IssueFence} event. Details of these are straightforward and are ommited for brevity.

%
%






%

\subsection{Modelling the HSA Atomics}

Again, using the definitions of \emph{po}, \emph{co} and \emph{hb} from the HSA specification, we are now able to model the HSA atomics.

First, we define the atomic LOAD, \emph{SCACQLOAD}, with \emph{acquire} semantics and the atomic STORE, \emph{SCRELSTORE} with \emph{release} semantics, and differentiate them from the \emph{ordinary} LOADs and STOREs.

As we did to model fences, above, we then refine the events of the generic model to establish this differentiation, imposing the ordering defined in the HSA specification in the following, refined\emph{Issue} events,
\emph{IssueAtomicSCACQLoad, IssueLoad, IssueAtomicSCRELStore, IssueStore}
and their corresponding \emph{Observe} events.

\subsection{Introducing the Register File}

We have a model which implements fences and atomics, which can be refined to introduce the register file as an Event-B variable, \emph{rf}

\begin{description}
		\nItem{ inv1 }{ rf \in  MASTER \pfun  (REG \tfun  VALUE) }
\end{description}

When a LOAD, \emph{l}, is observed by a master, \emph{m}, the register \emph{r} associated with \emph{l}, \emph{rl}, takes the last observed value for that LOAD location, \emph{addr(l)}.  For instance, in the event \emph{ObserveLoadHappensBeforewithFence} described above, an extra action, \emph{act3} is added.

\begin{description}
			\nItemX{ act3 }{ rf(m) :=  rf(m) \ovl  \{ r(l) \mapsto  (lov(m))(addr(l))\}  }
\end{description}

\section{Validating the Memory Model}

\subsection{A Notation for Representing Litmus Tests}


The memory model we have developed represents all the legal interleavings of an arbitrary set of programs running on an arbitrary set of masters. We now wish to constrain the model to represent a litmus test  -  all possible interleavings for a fixed set of masters each running a specific, fixed program thread.

\begin{figure}[!htb]
  \centering
  \includegraphics[width=9.5cm, height=4.5cm]{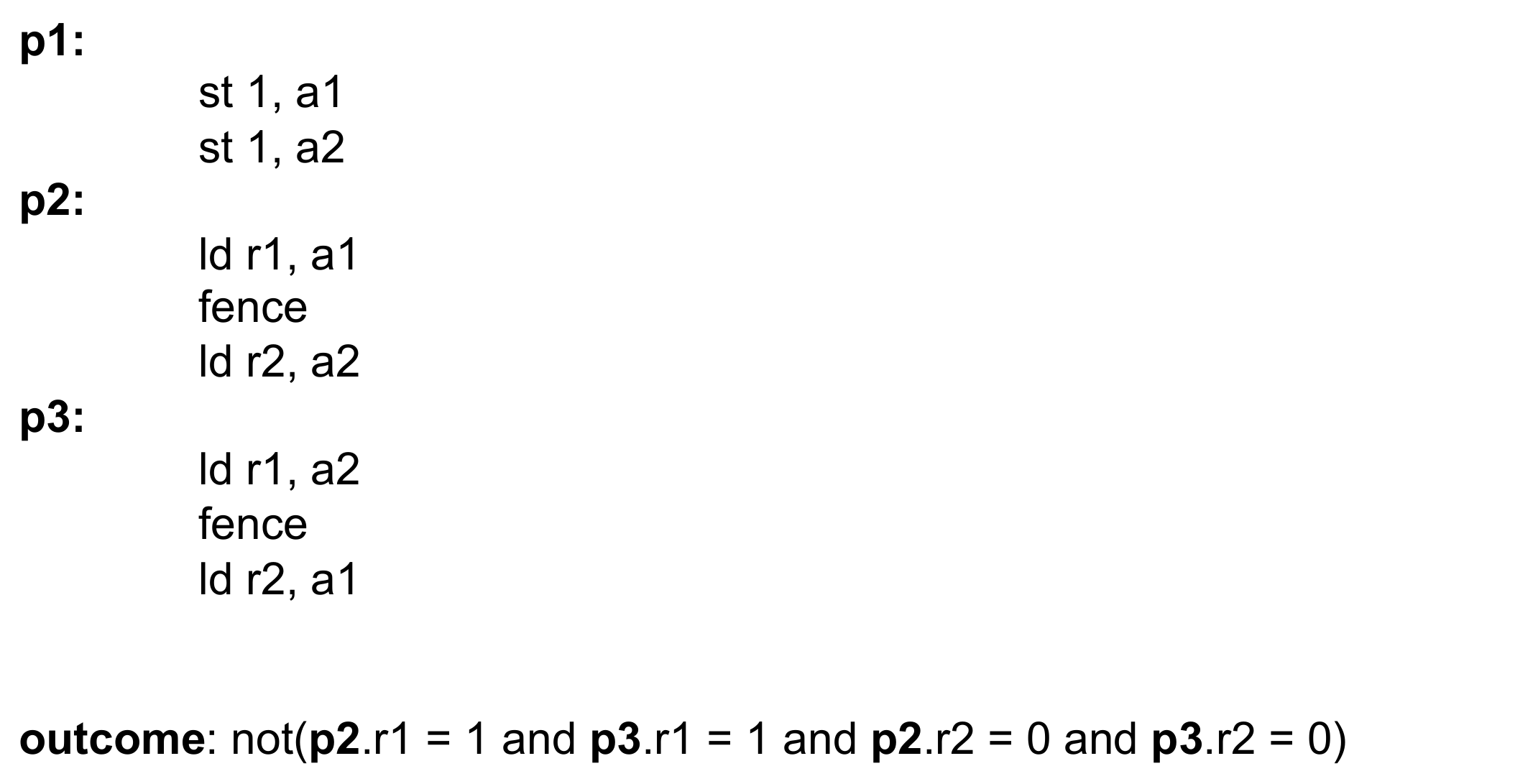}
  \caption{Litmus Test in Tool-supported Notation}
  \label{Figure:figNotation02} 
\end{figure}

To make it easier to verify litmus tests, we introduce a tool-supported notation with which the litmus test and the invariant can be represented and an Event-B \emph{context} and \emph{invariant} is generated automatically from this description.
An example of a litmus test in our notation as shown  in Figure \ref{Figure:figNotation02}.

\subsection{The Litmus Test Context}

We constrain the memory model by introducing axioms which specify the litmus test instructions, where for  instance \emph{I11} is a STORE of the value \emph{1} to the address \emph{a1}, and the program threads that will run those instructions.


\begin{description}
\nItem{ axm10 }{ P1 = \{ 1 \mapsto  I11, 2 \mapsto I12\}  }	\nItem{ axm11 }{ P2 = \{ 1 \mapsto  I21, 2 \mapsto  I22, 3 \mapsto  I23\}  }		\nItem{ axm12 }{ P3 = \{ 1 \mapsto  I31, 2 \mapsto  I32, 3 \mapsto  I33\}  }
\end{description}

We then associate each thread with the appropriate master.

\begin{description}
\nItem{ axm19 }{ PROGRAM = \{ M1 \mapsto  P1, M2 \mapsto  P2, M3 \mapsto  P3\}  }
\end{description}

\subsection{Verifying the Litmus Test Outcome}

An \emph{outcome} statement may be associated with the Litmus Test, which is represented as a boolean expression in terms of the values in the registers of the masters. We specify \emph{outcome} to mean
\emph{the values in the registers when all litmus test loads have been observed.}
An invariant is generated that we verify using the ProB model checker.  In the case of this litmus test, the invariant is
\begin{description}
		\nItem{ inv1 }{ \{ I21, I23, I31, I33\}  \subseteq  observed \limp \\
		\lnot ((rf(M2))(R1) = V1 \land  (rf(M3))(R1) = V1  \land \\
		((rf(M2))(R2) = V0 \land  (rf(M3))(R2) = V0)) }\\
	\end{description}
	
If the loads, \emph{I21, I23, I31, I33} issued by the masters \emph{M2} and \emph{M3} have been observed, then the registers associated with the masters cannot contain the prohibited outcome.  The values \emph{V0} and \emph{V1} represent logic \emph{0} and \emph{1} respectively.

\subsection{Covering the Allowable Register Value Combinations}

The facility to specify and verify the illegal litmus test outcomes formally is valuable, but to ensure thorough verification, it is necessary to go further.

\begin{itemize}
\item To ensure that all the \emph{legal outcomes} of the litmus test are \emph{reachable}.
\item To ensure that the set of litmus tests \emph{cover} the model functionality.
\end{itemize}



We introduce a further refinement of the memory model to introduce a \emph{functional coverage metric} for the litmus test.

The relation \emph{coverage}
	\begin{description}
		\nItem{ inv1 }{ coverage \in  REGCOVER \rel  REGCOVER }
	\end{description}
represents the combinations of register values that are reached by the litmus test, where \emph{REGCOVER} is defined in extended context thus:
	\begin{description}
		\nItem{ axm1 }{ C0 = \{ R1 \mapsto  V0, R2 \mapsto  V0\}  }		\nItem{ axm2 }{ C1 = \{ R1 \mapsto  V0, R2 \mapsto  V1\}  }		\nItem{ axm3 }{ C2 = \{ R1 \mapsto  V1, R2 \mapsto  V0\}  }		\nItem{ axm4 }{ C3 = \{ R1 \mapsto  V1, R2 \mapsto  V1\}  }		\nItem{ axm5 }{ REGCOVER = \{ C0, C1, C2, C3\}  }	\end{description}
%
Each constant C\emph{n} represents a register combination value for a master and \emph{REGCOVER} is the set of all register combinations for a master.  The \emph{coverage} relation is initialised to the empty set.
The refinement then introduces a new event \emph{CoverRegisterValues}

\begin{description}
	\EVT {CoverRegisterValues}
		\begin{description}
		\AnyPrm
			\begin{description}
			\Item{rm2, rm3 }
			\end{description}
		\WhereGrd
			\begin{description}
			\nItem{ grd1 }{ rm2 \in  REGCOVER }
			\nItem{ grd2 }{ rm3 \in  REGCOVER }
			\nItem{ grd3 }{ rf(M2) = rm2 }
			\nItem{ grd4 }{ rf(M3) = rm3 }
			\nItem{ grd5 }{ rm2 \mapsto  rm3 \notin  coverage }
			\nItem{ grd6 }{ \{ I21, I23, I31, I33\}  \subseteq  observed }
			\end{description}
		\ThenAct
			\begin{description}
			\nItem{ act1 }{ coverage :=  coverage \bunion  \{ rm3 \mapsto  rm4\}  }
			\end{description}
		\EndAct
		\end{description}
\end{description}

This event is enabled when all the LOADs have been observed (guard \emph{grd6})  and when this combination of the master register values has not already been covered (guard \emph{grd5}).  The combination is then added to the coverage relation.

We can now re-run the model checker exhaustively to show, not only that illegal register combinations are never reached but also that all legal combinations are reachable.  The ProB model checker has a coverage option that now allows us to display the values of the relation \emph{coverage} that are covered.


Coverage point 15 represents the initial value of the coverage relation.  Note that otherwise all 15 possible combinations have been covered and the illegal combination is not covered.


\subsection{Generating a Test for each Coverage Point}

Now that we are sure that each legal combination of registers is reachable, we wish to create a regression test for each of the 15 reachable combinations.  Creating the tests manually is time consuming.  ProB provides a \emph{model-checking-based} test generation facility which can be used to generate these tests automatically.

First, we specify a predicate to define the target state.  For instance, if we wish to show that all the registers of both masters can take the value 0 when all the LOADs have been observed, then we specify.
\\

${I21, I23, I31, I33} \subseteq observed => C0 \mapsto C0 : coverage$
\\
\\
Second, we specify the event or events which we wish to be covered by the test. In this case, we just specify the \emph{CoverRegisterValues} event to guide the search.

ProB generates the tests as an HTML file, which can be translated to suit the tool chain.
One of the tests that ProB generates is shown in Figure \ref{Figure:figTGEN0000} below.  The test describes a sequence of six \emph{Issue} instructions which results in the expected register value combination to be recorded by the event \emph{CoverRegisterValues}.

\begin{figure}[!htb]
  \centering
\begin{verbatim}
￼Test_case_id = 1
IssueLoad(I21, P2); 
IssueLoad(I31, P3); 
IssueFence(I22, P2); 
IssueLoad(I23, P2); 
IssueFence(I32, P3); 
IssueLoad(I33, P3);
CoverRegisterValues(rm2(R1->0, R2->0), 
                    rm3(R1->0, R2->0) ); 
\end{verbatim}
  \caption{Generated Test}
  \label{Figure:figTGEN0000} 
\end{figure}

\subsection{Measuring the Coverage of a set of Litmus Tests}

Since the litmus tests each target  a particular aspect of the weak memory model, an individual litmus test will not cover all of the events of the model.  What we do want to know, however, is that all events of the model are covered by the set of litmus tests.  In other words, that the litmus tests \emph{fully cover} the model functionality.

We return to the tests generated above and now, instead of specifying the \emph{CoverRegisterValues} event, we specify \emph{all} the events that we expect to be covered by the test.  We want to verify that these and only these events are covered by the litmus test. ProB will now generate several tests for the required outcome which cover the events specified.

\subsection{Generating additional Litmus Tests}

Once we have completed the model validation and have a set of litmus tests which cover the model, we relax the constraints on our formalised litmus tests so that each model represents a class of programs with a non-deterministic mix of instructions (and the same set of masters, registers and address space as the associated litmus).  Using the ProB test generator, we are able to generate automatically a wide range of tests, with expected register values, which can be used for regression testing of the HSA-compliant platform itself.

\section{Related Work and Conclusion}

An overview of memory ordering and barriers in modern microprocessors is presented in \cite{mckenney2005memory1} and\cite{mckenney2005memory2}.
The release consistency model is described in \cite{gharachorloo1990memory}.
The notion of using thread-local re-ordering and store atomicity, together with the definition of an abstract architecture to support weakly consistent memory specification is developed in \cite{Shen99commit-reconcilefences} and \cite{arvind2006memory}.  The tutorial 
\cite{maranget2012tutorial}, for ARM and POWER memory models, builds on the notion of an abstract machine architecture to define a \emph{storage subsystem} which has general applicability for weakly consistent memory specification.  \cite{pratt1986modeling} introduces the notion of modelling concurrency with partial orders.   \cite{mador2012axiomatic} presents an axiomatic approach to POWER  memory modelling and \cite{alglave2010fences}  presents a class of relaxed memory models which are parameterised to support different local re-orderings and store atomicity relaxation. \cite{chong2008reasoning} provides a clear description of the ARM memory model and the HSA initiative to harmonise heterogenous computing concepts \cite{HSAPSAS}  has resulted in a natural language specification of fences and atomics which provides considerable clarity for programmers, model and system developers.

In \cite{alglave2014herding}, an \emph{axiomatic} approach is used where the memory model is represented as a directed graph, the nodes representing the memory accesses and the arcs the ordering between those accesses. Our Event-B approach is based on the computation model of \emph{guarded atomic actions}.  Memory access ordering is specified \emph{axiomatically} as guards on events, which model the memory accesses themselves, for instance \emph{IssueRead}, \emph{ObserveWrite}. If more than one event is enabled, then a choice is made non-deterministically, naturally representing the non-determinism of the memory model. The Event-B model is operational in the sense that it is based on the highly abstract memory architecture described in \cite{Shen99commit-reconcilefences}.  We begin with a generic model to represent the ordering at a high level of abstraction and then use formal refinement to introduce, in steps, as much detail as is necessary to represent a concrete, concurrent program, where the state of the registers and the values observed by each of the masters are updated by the event actions.  The concrete model can therefore be mapped directly to a Transaction-Level Model (TLM) for efficient system simulation.  \cite{alglave2014herding} also presents a tool for simulation and testing of memory models. Using ProB we can simulate and test our Event-B models at each level of abstraction as well as generate coverage-driven tests for regression testing of hardware and software systems.

We believe our methodology of refinement based formal modelling, baseline compliance testing of the formal model and coverage driven test extraction using a single language of Event B language and the Rodin tool is a completely new way of addressing a profoundly important challenge facing the design and verification of low-power of multi-core systems.
It provides a hierarchy of models  where the complexity of modelling is managed by refinement leading to clarity in understanding and precision of formal modelling. 
Using the same Event B model on which we have shown that refinements are provably correct, we can test it for litmus test compliance against published HSA litmus tests. The key here is that testing for compliance is embedded within the Rodin framework where proof based refinement is done; leading to a very tight integration of testing and proofs.
By relaxing the constraints of HSA litmus tests we are able to explore a much bigger space of tests and by a natural extension of the core Event B model of the HSA by a coverage event we are able to generate  a bigger set of regression tests that satisfy the coverage model. This is again done on the same core Event B model on which proof based refinements were done.
The tests generated can be used in the development and verification of the HSA-compliant platform itself.
\\




\section*{Acknowledgment}
The authors would like to thank Jason Meredith, Mark Landers and James Aldis for several discussions during the development of this work. The authors would also like to thank Imagination Technologies for their time and support throughout this work.



%
\bibliographystyle{IEEEtran}
\bibliography{IEEEabrv,weakmemory.bib}




\end{document}